\documentclass[singlecolumn,showpacs,preprintnumbers,amsmath,amssymb]{revtex4}
\usepackage{graphicx}
\usepackage{dcolumn}
\usepackage{bm}

\begin{document}

\title{Study of fusion dynamics using Skyrme energy density formalism with different surface corrections}

\author{Ishwar Dutt}%
 \email{idsharma.pu@gmail.com}
\affiliation{%
Department of Physics, Panjab University, Chandigarh -160014, (India)\\
}%

\author{Narinder K. Dhiman}
\affiliation{
Govt. Sr. Sec. School, Summer Hill, Shimla -171005, (India)\\
}%


\date{\today}

\begin{abstract}
Within the framework of Skyrme energy density formalism, we
investigate the role of surface corrections on the fusion of
colliding nuclei. For this, the coefficient of surface correction
was varied between 1/36 and 4/36, and its impact was studied on
about 180 reactions. Our detailed investigations indicate a linear
relationship between the fusion barrier heights and strength of
the surface corrections. Our analysis of the fusion barriers
advocate the strength of surface correction of 1/36.
\end{abstract}


\maketitle

\section{\label{intro}Introduction}
 The usefulness of the Skyrme energy density formalism (SEDF) in
understanding the fusion dynamics has been widely
accepted~\cite{rkp1,wang06,shen09}. Several ion-ion potential
models and parametrizations have been suggested based on the same
formalism~\cite{rkp1,wang06,shen09}. Interestingly, along with
SEDF, many other models/formalisms based on either
microscopic/macroscopic or phenomenological picture are found  to
reproduce  the fusion barrier heights within $\pm10\%$. In a
recent study~\cite{id}, we employed as many as 16 versions of
different proximity potentials and found that one can reproduce
the barrier heights across the periodic table within $\pm8\%$.
Even in the last few years, many sets of the Skyrme forces that
reproduce the ground states properties of large number of nuclei
have also been proposed~\cite{rkp1,wang06,shen09}. Within the same
SEDF, the use of different Skyrme forces can yield difference  of
$\pm10\%$ for fusion barriers. The same Skyrme forces have also
been used at intermediate energies to investigate many rare
phenomena~\cite{sood}. The strength and form of nuclear potential
is also very important in the study of cluster decay~\cite{rkg}.
\par
 Generally, SEDF consists of Hamiltonian, that depends on the nucleonic
density, kinetic energy density as well as on the spin
density~\cite{rkp1}. Among all these, the form and strength of the
kinetic energy density has always controversial in  the
literature~\cite{graf80}. Mostly reported works used
approximations  based on the Thomas-Fermi
approximation~\cite{graf80,ms74,Hilf66,bathe68}.
\par
In the literature, additional correction in term of gradient term
has been suggested over and above the Thomas-Fermi approximation
~\cite{graf80,ms74,Hilf66,bathe68,von35}. Though all studies
advocate the inclusion of this term, its strength has yet not been
resolved and many different strengths are available in the
literature~\cite{graf80,von35,gupta85}. We plan to explore the
role of this surface correction term in heavy-ion collisions via
fusion process and want to understand whether one can narrow down
the choice of this parameter or not. We shall also attempt to
present a simple parametrization of this term for fusion barriers.
This aim is achieved by employing SEDF within the proximity
concept discussed in brief in section \ref{model}. The results are
presented in section \ref{result} and a brief summary is presented
in section \ref{summary}.
\par
\section{\label{model} Skyrme Energy Density Formalism:}
In the Skyrme energy density formalism~\cite{rkp1}, the nuclear
part of the interaction potential $V_{N}(r)$ is calculated as a
difference of the energy expectation value at a distance $r$ and
at infinity (i.e. at $r=~ \infty $):
\begin{equation}
V_{N}\left(r \right)  = E\left(r \right)- E\left(\infty  \right),
\label{eq:1}
\end{equation}
with
\begin{equation}
E = \int H \left(\vec{r} \right)\vec{dr}. \label{eq:2}
\end{equation}
In our formalism, the energy density functional $H \left(\vec{r}
\right)$ read as;
\begin{eqnarray}
H(\rho,\tau,\vec{J}) & = & \frac{\hbar^2}{2m} \tau +\frac{1}{2}t_0
[(1+
\frac{1}{2}x_0)\rho^2-(x_0+\frac{1}{2})(\rho_n^2+\rho_p^2)]+\frac{1}{4}(
t_1+t_2)\rho\tau \nonumber\\
& & +\frac{1}{8}(t_2-t_1)(\rho_n \tau_n+\rho_p \tau_p)
+\frac{1}{16} (t_2-3t_1) \rho \nabla^2 \rho \nonumber \\ & &
+\frac{1}{32} (3t_1+t_2) (\rho_n \nabla^2 \rho_n+\rho_p \nabla^2
\rho_p) + \frac{1}{4}t_3 \rho_n \rho_p \rho \nonumber \\ & &
-\frac {1}{2} W_0(\rho \vec{\nabla}\cdot\vec{J} +\rho_n
\vec{\nabla}\cdot\vec{J}_n+ \rho_p \vec{\nabla} \cdot\vec{J}_p) .
\label{eq:3}
\end{eqnarray}
Here $\rho$ is the nucleon density taken to be two-parameter Fermi
density and $\vec{J}$ is the spin density which was generalized by
Puri et al., for spin-unsaturated nuclei~\cite{rkp1}. The
remaining term is the kinetic energy density $\tau$. The most
widely used Skyrme force type SIII~\cite{bn75} is used in the
present analysis.

\par
  The evaluation of kinetic energy density term was done within the
Thomas-Fermi (TF) approximation which is a well known alternative
to the Hartree-Fock (HF) method. As shown by Myers and
\'Swiatecki~\cite{ms74} and Hilf and
S$\ddot{\rm{u}}$ssman~\cite{Hilf66}, the kinetic energy density
$\tau$ can be separated into volume term $\tau_{0}$ and surface
term $\tau_{\lambda}$ plus reminder. In the first order
approximation, one can limit to $\tau_{0}$=
$\frac{3}{5}\left(\frac{3}{2}\pi^{2}\right)^{\frac{2}{3}}\rho^\frac{5}{3}$
term only. However, due to the serious drawback in explaining the
nuclear surface as well as the densities, a demand of further
correction to this term was realized. Bethe and
Brueckner~\cite{bathe68} proposed that this difficulty can be
overcome by adding some gradient term and correction proportional
to $\frac{\left(\vec{\nabla} \rho\right)^2}{\rho}$; derived long
ago by von-Weizs$\ddot{\rm{a}}$cker~\cite{von35}. Alternative to
this is the gradient extension method that also yields similar
results. The coefficient in the above term was first thought to be
close to 1/4 ( = 9/36)~\cite{graf80}, later on, it was  found that
this strength does not give proper results for kinetic energy
density. Following this, several authors suggested different
values for surface
correction~\cite{graf80,ms74,Hilf66,bathe68,von35,gupta85} that
ranges from 1/36 to 9/36. The kinetic energy density can therefore
be written as;
\begin{equation}
\tau =  \tau_{0}+ \tau_{\lambda}= \tau_{0} + \lambda
\frac{\left(\vec{\nabla} \rho\right)^2}{\rho}, \label{eq:6}
\end{equation}
where $\lambda$ is an adjustable  parameter. As discussed by
Gr$\ddot{\rm{a}}$f~\cite{graf80},  its value is supposed to be
between $\frac{1}{36}$ and $\frac{9}{36}$. In last few years,
large number of theoretical models however used the value of
$\lambda=\frac{1}{36}$~\cite{wang06,gupta85,breck85}. One is
wondering how $\lambda$  can alter the results of fusion barriers
 and further can we understand its effect in terms of  some parameterized form or
 not. It is worth mentioning that different strengths of $\lambda$ can be thought to be close to
 Yukawa term used in addition to the Skyrme forces at intermediate energies~\cite{sood}.
 Therefore, this term is very important for reproducing the
surface properties of nuclei as well as for multifragmentaion.  We
shall calculate the ion-ion potential using the above formalism
for different values of $\lambda$ and then extract the fusion
barrier heights and positions.


%
\section{\label{result}Results and Discussions}
In the present study, as many as 180 reactions involving even-even
masses between 24 and 246 are taken for analysis. As noted,
symmetric  as well as asymmetric colliding nuclei  have been taken
into account. We calculated the ion-ion potential for all 180
reactions using different surface strengths. The  heights and
positions of the barriers were calculated using conditions:
 \begin{equation}
\frac{dV_T(r)}{dr}|_{r=R_{B}} = 0,~~ \rm{and} ~~
\frac{d^{2}V_T(r)}{dr^{2}}|_{r=R_{B}} \leq 0. \label{eq:8}
\end{equation}
For the present analysis, we took  $\lambda$ = $\frac{1}{36}$,
$\frac{2}{36}$, $\frac{3}{36}$, and $\frac{4}{36}$. The barrier
height is denoted by $V_{B}$ and its corresponding position is
marked as $R_{B}$. We shall also attempt to parameterize the
fusion barriers, thus, obtained and will present unique
correlation between different values of $\lambda$.
\par
 In Fig. 1, we
display the percentage difference of the barrier heights and
positions calculated using different values of $\lambda$ (between
$\frac{1}{36}$ and $\frac{4}{36}$) over experimental values
defined as:
\begin{equation}
\Delta V_{B}~\%= \frac{V_{B}^{theor} -
V_{B}^{expt}}{V_{B}^{expt}}\times 100, \label{eq:9}
\end{equation}
 as a function of $Z_{1}Z_{2}$. The experimental data are
 taken from Refs.~\cite{rkp1,Vaz81,cm81,gary82,sb81,Aguilera86,stefani09,trotta01,Ichikawa09,Sonzogni98,vinod96}.
 We see that different values of $\lambda$ can explain  barrier
 heights within $\pm~10\%$ of the experimental values. Whereas, due to large uncertainty in
 the measurement of barrier positions the deviation in some cases can be
 quite large~$\pm~30\%$. These data have been of controversy in the literature~\cite{rkp1}.
 Due to different experimental setups, measurements  do not
 yield few barrier positions as per known trend. A more careful
 look reveals that fusion barrier heights calculated with $\lambda$ = $\frac{1}{36}$ yield better results on
 average as compared to other values of surface correction used. From this figure,
 which is spanned over the entire periodic table, it is evident that
$\lambda$ = $\frac{1}{36}$ may be preferred  as the coefficient of
the gradient term in the kinetic energy density correction. This
finding is in agreement with other large number of studies based
on the structural aspect of heavy-ion
reactions~\cite{wang06,gupta85,breck85}. However slight deviations
in some cases may be due to the experimental uncertainties
reported in many papers~\cite{trotta01} or it may also be due to
the influence of additional higher order correction terms that we
have not taken into account in the present calculations.
\par
 We shall now attempt to parametrize the fusion barrier heights
 and positions using different $\lambda$ - values in terms of
 charges and masses of the colliding pair.
In Fig. 2, we display the barrier positions $R_{B}$ as a function
of $(A_{1}^{1/3}+A_{2}^{1/3})$ and $V_{B}$ as a function of
$\frac{Z_1 Z_2e^{2}}{R_B^{anal}}(1-\frac{1}{R_B^{anal}})$; where
$R_{B}^{anal}$
 is the analytical barrier positions obtained after parametrization
 have been performed over $R_{B}$ results.  When we increase the
 value of $\lambda$  from 0 to $\frac{1}{36}$, $\frac{2}{36}$,
 $\frac{3}{36}$, and $\frac{4}{36}$, a monotonous
 increase in the nuclear part of the potential results in the reduction of the height of the barrier and therefore,
 the barrier positions are pushed outward.
  The addition of the strength $\lambda$ increases the attractive part of the
 interaction potential. In  other words, one can counterbalance the repulsive
 Coulomb potential at larger distances, therefore, pushing
 the barrier outwards. As a result, net decrease in the barrier height occurs. In Fig. 2, we display the results with
$\lambda$ = $\frac{1}{36}$, = $\frac{3}{36}$, and =
$\frac{4}{36}$. The left part is for the barrier positions whereas
right part is for the barrier heights. The barrier heights and
positions are parametrized in terms of the following relations:

\begin{equation}
R_{B}^{anal} =a^{'}+b^{'}X_{1}; ~~~~~~~~~~~{\rm and}
~~~~~~~~~~~V_{B}^{anal} =c^{'}X_{2},\label{eq:10}
\end{equation}
where $X_{1}=(A_{1}^{1/3}+A_{2}^{1/3})$, and $X_{2}=\frac{Z_1
Z_2e^{2}}{R_B^{anal}}(1-\frac{1}{R_B^{anal}})$. Here $a^{'}$,
$b^{'}$, and $c^{'}$ are the constants displayed in the figure. We
see a linear increase in the barrier heights and positions with
the masses of the colliding nuclei. This is in agreement with many
previous studies~\cite{cw76}. With the increase in the strength of
$\lambda$, a monotonous  decrease in the barrier height and
increase in  the barrier positions can be seen. In all  cases, a
linear fit can explain the effect of $\lambda$ on the barrier
positions very well.
\par
In Fig. 3, we display the percentage difference between the
analytical and exact values as a function of $Z_{1}Z_{2}$. Very
interestingly, we note that we can reproduce the actual barriers
(heights  as well as positions) within $\pm ~2.5\%$ for  all
values of $\lambda$. This introduces great simplification in
calculating the barrier positions using different strengths
$\lambda$. It would be of further interest to understand  how
different values of $\lambda$ affects the fusion barriers in
different reacting nuclei.
\par
In Fig. 4, we display the parametrized fusion barrier heights
$V_{B}^{anal}$, as a  function of the different values of
$\frac{\lambda}{36}$ for the reactions of $^{24}Mg+^{24}Mg$,
$^{28}Si+^{28}Si$, $^{48}Ca+^{48}Ca$, $^{64}Ni+^{64}Ni$,
$^{40}Ca+^{50}Ti$, and $^{48}Ti+^{58}Ni$.  Interestingly,  we note
that the barrier heights reduce systematically as one move from
$\lambda~=~0.0$ to $\lambda~=~\frac{4}{36}$ values. This reduction
can be further quantify by the relation:
\begin{equation}
V_{B}^{anal} =\alpha(1-\delta\frac{\lambda}{36}). \label{eq:10}
\end{equation}
Here $\alpha$ and $\delta$ are the constants whose values depend
upon the colliding pair (see Fig. 4). Therefore, decrease in the
barrier height with $\lambda$ is on expected lines, a linear
reduction is of particular interest.
\par
Finally, in Fig. 5, we display the fusion cross sections
$\sigma_{fus}$ calculated  using the formula of Wong~\cite{wg72},
as  a function of the center-of-mass energy, $E_{c.m.}$, for the
reactions of $^{24}Mg+^{24}Mg$~\cite{cm81,gary82},
$^{28}Si+^{28}Si$~\cite{gary82,sb81,Aguilera86},
$^{48}Ca+^{48}Ca$~\cite{stefani09,trotta01},
$^{64}Ni+^{64}Ni$~\cite{Ichikawa09},
$^{40}Ca+^{50}Ti$~\cite{vinod96}, and
$^{48}Ti+^{58}Ni$~\cite{Sonzogni98}. We see that no particular
value of $\lambda$ explains  the fusion cross sections. In few
cases, however, $\lambda=\frac{1}{36}$ yields better results
whereas in other cases, higher values of $\lambda$ has better
edge.
\section{\label{summary}Summary}
In the present study, we investigated the effect of surface
corrections on the fusion process.  Our finding over 180 reactions
reveals that $\lambda=\frac{1}{36}$ can be a better choice for the
surface correction and it is in agreement with earlier attempts.
We also obtained parameterized form of the fusion barrier heights
and positions for different strength of $\lambda$. We further
found that the fusion barrier heights depend linearly on the
strength of
the coefficient. \\

\newpage

\begin{figure}
\centering
\includegraphics* [scale=0.40] {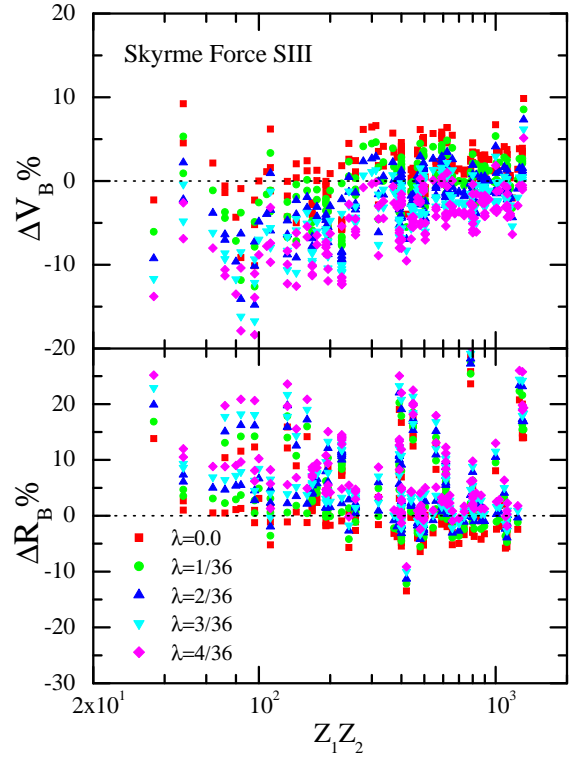}
 \caption {(Color online) The percentage difference
between the theoretical and experimental values as  a function of
$Z_{1}Z_{2}$ for different values of surface correction
$\lambda$.}
\end{figure}
\begin{figure}
\centering
\includegraphics* [scale=0.40] {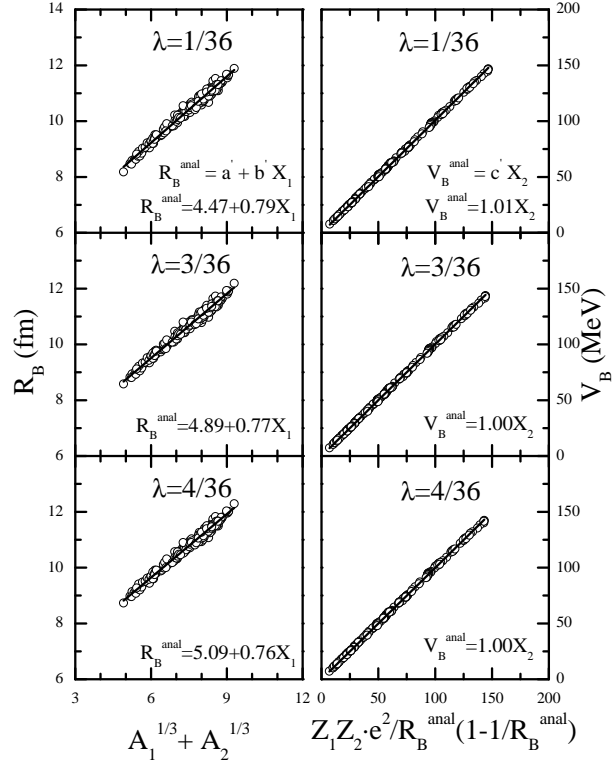}
\caption {The barrier positions $R_{B}$ (left side) and barrier
heights $V_{B}$ (right side) as a function of
$A_{1}^{1/3}+A_{2}^{1/3}$ and  $\frac{Z_1
Z_2e^{2}}{R_B^{anal}}(1-\frac{1}{R_B^{anal}})$, respectively for
$\lambda=1/36$, $\lambda=3/36$, and $\lambda=4/36$. The solid
lines represents the straight line linear fit
 over the points.}
\end{figure}
\begin{figure}
\centering
\includegraphics* [scale=0.40] {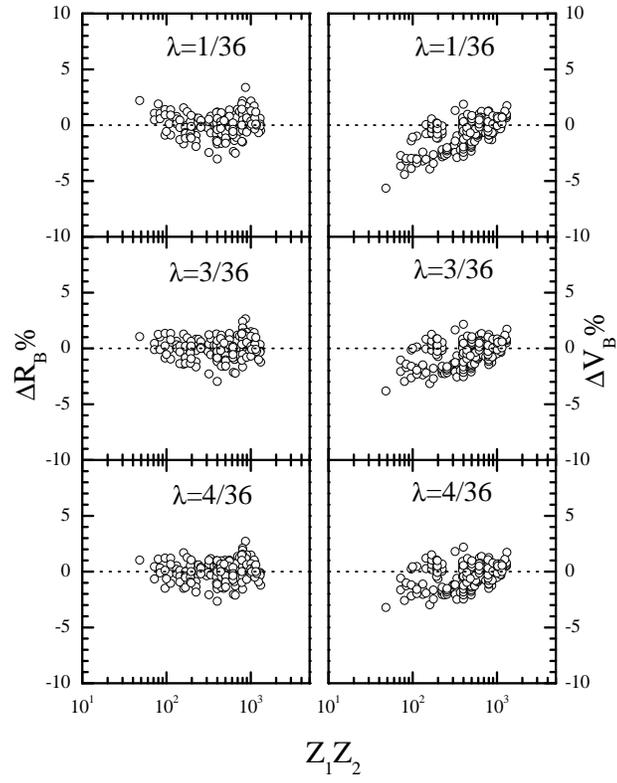}
\caption {The percentage difference between the analytical
 and theoretical values as  a function of $Z_{1}Z_{2}$
for different values of surface correction $\lambda$.}\label{fig3}
\end{figure}
\begin{figure}[!t]
\begin{center}
\includegraphics*[scale=0.4] {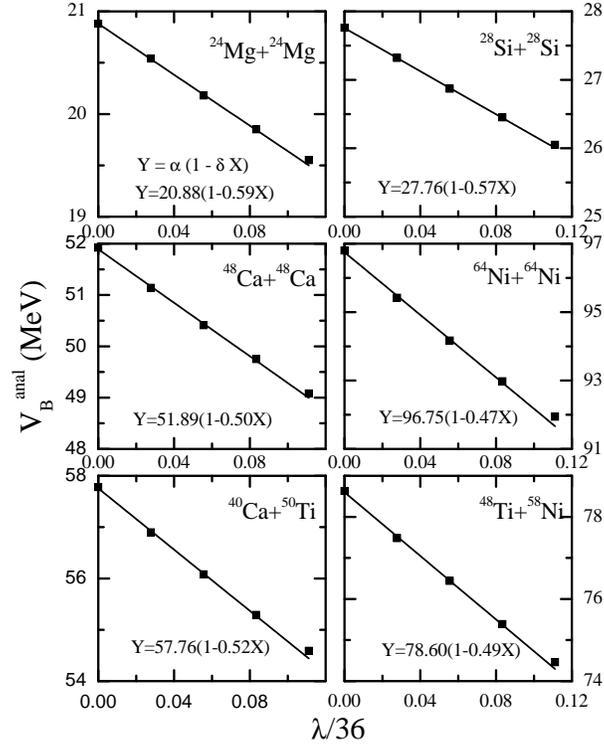}
 \caption {The analytical barrier heights
$V_{B}^{anal}$ as a function of $\lambda/36$ for the reactions of
$^{24}Mg+^{24}Mg$, $^{28}Si+^{28}Si$, $^{48}Ca+^{48}Ca$,
$^{64}Ni+^{64}Ni$, $^{40}Ca+^{50}Ti$, and $^{48}Ti+^{58}Ni$. The
solid lines represent the straight-line linear fit
 over the points.}
\end{center}
\end{figure}
\begin{figure}[!t]
\centering
\includegraphics* [scale=0.42]{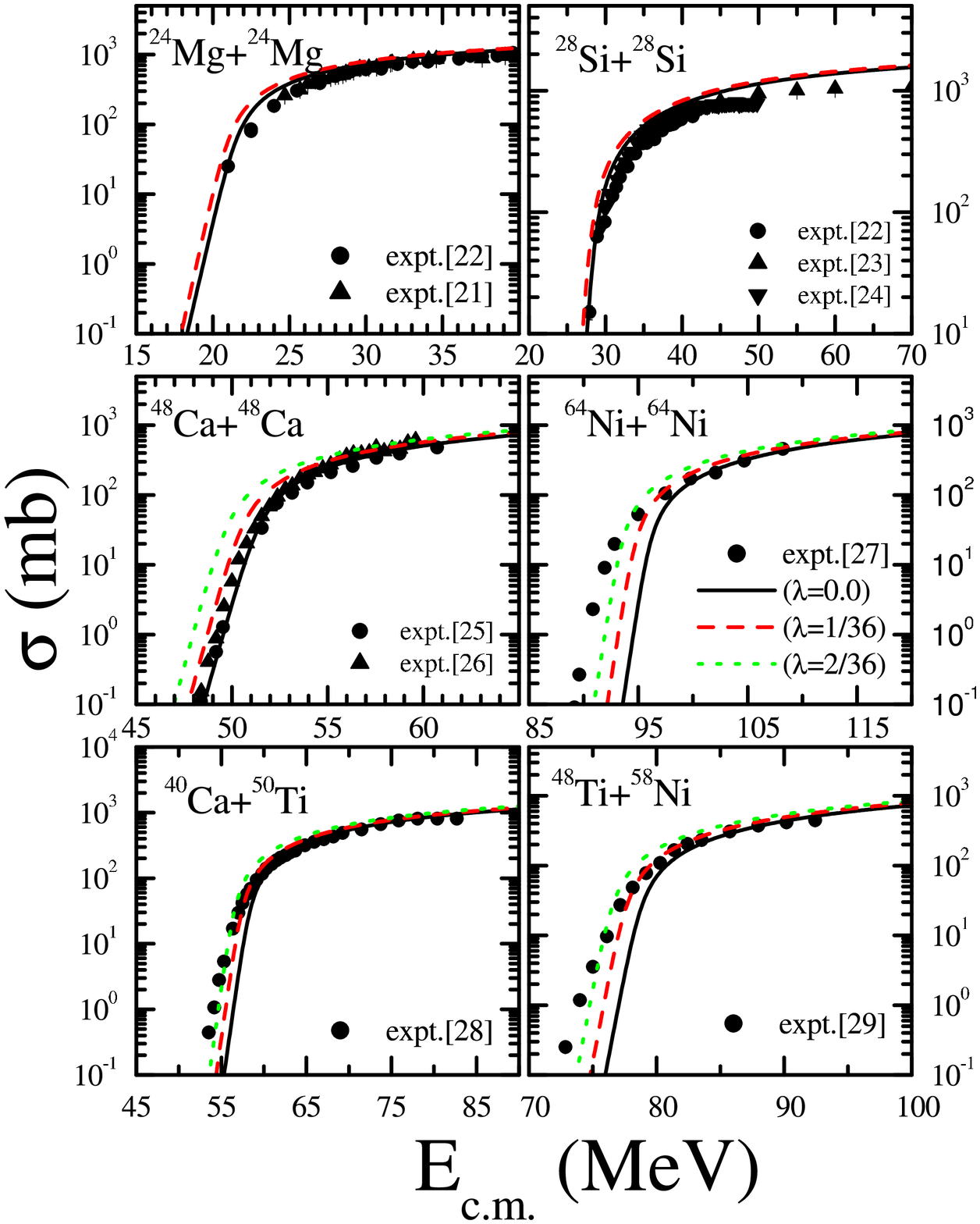}
\caption {(Color online) The fusion cross sections $\sigma_{fus}$
(mb) as a function of the center-of-mass energy $E_{c.m.}$ for the
reactions of $^{24}Mg+^{24}Mg$, $^{28}Si+^{28}Si$,
$^{48}Ca+^{48}Ca$, $^{64}Ni+^{64}Ni$, $^{40}Ca+^{50}Ti$, and
$^{48}Ti+^{58}Ni$ using different values of surface correction
$\lambda $.}\label{fig4}
\end{figure}


\section*{References}
\begin{thebibliography}{0}
\bibitem{rkp1} R. K. Puri \emph{et al.},  Eur. Phys. J. A \textbf{23},
429 (2005); R. Arora \emph {et al.},  \emph {ibid.} \textbf{8},
103 (2000); R. K. Puri \emph {et al.},  \emph {ibid.} \textbf{3},
277 (1998); R. K. Puri \emph {et al.}, Phys. Rev. C \textbf{43},
315 (1991);  \emph {ibid.} \textbf{45}, 1837 (1992).
\bibitem{wang06} A. Dobrowolski, K. Pomorski, and J. Bartel, Nucl. Phys. A \textbf{729}, 713 (2003);
V. I. Zagrebaev,  \emph{ibid.} \textbf{734}, 164 (2004);  N. Wang,
J. Li, E. Zhao, Phys. Rev. C \textbf{78}, 054607 (2008).
\bibitem{shen09} Q. Shen, Y. Han, H. Guo,  Phys. Rev. C \textbf{80}, 024604
 (2009); Z. Feng, G. Jin, F. Zhang, Nucl. Phys. A \textbf{802}, 91
 (2008).
\bibitem{id} I. Dutt and R. K. Puri  Phys. Rev. C
\textbf{81}, 044615 (2010); \textbf{81}, 064609 (2010).
\bibitem{sood} Y. K. Vermani  \emph{et al.}, J. Phys. G: Nucl. Part. Phys.
\textbf{36}, 105103 (2009); S. Kumar \emph{et al.}, Phys. Rev. C
\textbf{81}, 014601 (2010); \textbf{58}, 3494 (1998); \textbf{78},
064602 (2008); \textbf{81}, 014611 (2010); J. Singh \emph{et al.},
Phys. Rev. C \textbf{62}, 044617 (2000); J. Dhawan \emph{et al.},
\emph{ibid.} \textbf{75}, 057601 (2007); R. K. Puri \emph{et al.},
Phys. Rev. C \textbf{54}, R28 (1996); J. Comput. Phys.
\textbf{162}, 245 (2000);  Y. K. Vermani \emph{et al.},  J. Phys.
G: Nucl. Part. Phys. \textbf{37}, 015105 (2010); Europhys. lett.
\textbf{85}, 62001 (2009);  Phys. Rev. C \textbf{79}, 064613
(2009); A Sood \emph{et al.}, \emph{ibid.} \textbf{70}, 034611
(2004); A Sood \emph{et al.}, \emph{ibid.} \textbf{79}, 064618
(2009);

\bibitem{rkg} R. K. Gupta \emph {et al.},  Phys. Rev. C \textbf{47}, 561
(1993);  J. Phys. G: Nucl. Part. Phys. \textbf{18}, 1533 (1992);
S. S. Malik \emph {et al.}, Pramana J. Phys. \textbf{32}, 419
(1989); R. K. Puri \emph {et al.}, Europhys. Lett. \textbf{9}, 767
(1989).
\bibitem{graf80} H. Gr\"af,  Nucl. Phys. A \textbf{343}, 91
(1980).
\bibitem{ms74} W. D. Myers and W. J. \'Swiatecki,  Ann. of
Phys. \textbf{84}, 186 (1974).
\bibitem{Hilf66} E. Hilf and G. S\"ussmann,  Phys. Lett.
\textbf{21}, 654 (1966).
\bibitem{bathe68} H. A. Bethe, Phys. Rev. \textbf{167}, 879
(1968); K. A. Brueckner and R. Buchler,  Proc. lnt. Conf. on few
particle problems in the nuclear interaction, Los Angeles, USA,
1972, ed. 1. Slaus, S. A. Moszkowskt, R. P. Haddock and W.T.H. van
Oers (North-Holland, Amsterdam, 1972) p. 913.
\bibitem{von35} C. F. von Weizs\"acker,  Z. Phys. \textbf{96}, 431
(1935).
\bibitem{gupta85} P. Chattopadhyay and R. K. Gupta,  Phys. Rev. C \textbf{30}, 1191
(1984).
\bibitem{bn75}  M. Beiner, H. Flocard, N. Van Giai and P. Quentin, Nucl. Phys. A \textbf{238}, 29 (1975).
\bibitem{breck85} M. Brack \emph{et al.},  Phys. Rep. \textbf{123}, 275
(1985).
\bibitem{Vaz81} Z. H. Liu \emph{et al.},  Eur. Phys. J. A \textbf{26}, 73 (2005); J. Skalski, Phys. Rev. C \textbf{76}, 044603 (2007);
S. Mitsuoka \emph{et al.}, Phys. Rev. Lett. \textbf{99}, 182701
(2007).
\bibitem{cm81} C. M. Jachcinski \emph{et al.},  Phys. Rev. C \textbf{24}, 2070
(1981).
\bibitem{gary82} S. Gary and C. Volant,  Phys. Rev. C \textbf{25},
1877 (1982).
\bibitem{sb81} S. B. DiCenzo \emph{et al.},   Phys. Rev. C {\bf 23,}
2561 (1981).
\bibitem{Aguilera86} E. F. Aguilera \emph{et al.},   Phys. Rev. C \textbf{33},  1961 (1986).
\bibitem{stefani09} A. M. Stefanini \emph{et al.}, Phys Lett. B \textbf{679}, 95
(2009).
\bibitem{Ichikawa09} T. Ichikawa \emph{et al.},  Phys. Rev. Lett. \textbf{103},  202701 (2009).
\bibitem{Sonzogni98} A. A. Sonzogni \emph{et al.},  Phys. Rev. C \textbf{57},  722 (1998).
\bibitem{vinod96} A. M. Vinodkumar \emph{et al.}, Phys. Rev. C  \textbf{53},  803
(1996).\

\bibitem{trotta01} M. Trotta  \emph{et al.}, Phys. Rev. C \textbf{65}, 011601(R) (2001); H. A. Aljuwair  \emph{et
al.}, \emph{ibid.} \textbf{30}, 1223 (1984); L. C. Vaz \emph{et
al.},  Phys. Rep. \textbf{69}, 373 (1981); D. G. Kovar  \emph{et
al.}, Phys. Rev. C \textbf{20} 1305 (1979).
\bibitem{cw76} P. R. Christensen and A. Winther,  Phys. Lett. B \textbf{65}, 19 (1976); R. M. Anjos \emph{et al.},  \emph{ibid.} \textbf{534}, 45 (2002);
R. A. Broglia, A. Winther, Heavy–Ion Reactions Lecture Notes
(Redwood City, CA: Addison-Wesley)  p. 116 (1981).
\bibitem{wg72} C. Y. Wong,  Phys. Lett. B \textbf{42}, 186 (1972);  Phys. Rev. Lett. \textbf{31}, 766 (1973).

\end{thebibliography}
\end{document}